\documentstyle[11pt,fleqn]{article}  
\begin{document}        
\title{Possible Knot-type Time-dependent Quantum-mechanically Dynamical System } 
\author{Zotin K.-H. Chu}  
\date{3/F, 4, Alley 2, Road Xiushan, Leshanxinchun, Xujiahui 200030, China}
\maketitle
\begin{abstract}
We illustrate  a  possible traversing along the path of
trefoil-type and $8_{18}$ knots during  a specific time period by
considering a quantum-mechanic system which satisfies a specific
kind of phase dynamics of quantum mechanics. This result is
relevant to the composite particle
which is present in the initial or final configuration. 
%
\end{abstract}             %
\doublerulesep=6.8mm    %
\baselineskip=6.8mm
\bibliographystyle{plain}               %
\section{Introduction}
Solutions of the time-dependent Schr\"{o}dinger equation have been
of considerable interest since quantum mechanics was established
[1-5]. Driven by the need to solve theoretical and practical
problems throughout physical fields, such as atomic physics,
condensed matter physics, perturbative and nonperturbative methods
were proposed and developed. One approach : Dirac's perturbation
theory [1-3], being one of the most successful methods in dealing
with time-dependent quantum systems, caused seemingly unnecessary
concerns. \newline Meanwhile, knots themselves are macroscopic
physical phenomena in three-dimensional space, occurring in rope,
vines, telephone cords, polymer chains, DNA, certain species of
eel and many other places in the natural and man-made world [6-13]
(to have knots at all, that is configurations of a curve falling
into a certain class which cannot be transformed into other
classes (i.e. different knots, including no knot), one must have
infinite or closed curves [6]). For example, the static and
dynamic behavior of single polymer chains, such as DNA, and
multichain systems like gels and rubbers, is strongly influenced
by knots and permanent entanglements [6,12-13]. The study of
topological invariants of knots leads to relationships with
statistical mechanics and quantum physics [6-13]. This is a
remarkable and deep situation where the study of certain
(topological) aspects of the macroscopic world is entwined with
theories developed for the subtleties of the microscopic world.
\newline
In this short paper, we shall illustrate the possible link between
the study of time-dependent quantum system and the application of
the relevant phase-part result to the simple knot, like the
trefoil and $8_{18}$-type.
\section{Formulation}
According to the basic formalism of quantum mechanics, we can
formally construct a unitary operator to obtain the dynamical
wavefunction $\Psi (t)$ from the initial wavefunction $\Psi
(t_0)$, namely, we have
\begin{displaymath}
 \Psi (t)=\exp[\frac{-i}{\hbar}\int^t_{t_0} H(\tau) d\tau]
 \Psi(t_0),
\end{displaymath}
where $H(\tau)$ is the time-dependent Hamiltonian [1-3]. We first
discretize the entire time span from $t_0$ to $t$ into $N$
intervals as $\Delta t_1=t_1-t_0$, $\Delta t_2=t_2-t_1, \cdots$,
$\Delta t_N=t-t_{N-1}$. We thus have
\begin{equation}
\Psi (t)=\exp[\frac{-i}{\hbar}\int^t_{t_{N-1}} H(\tau) d\tau]
\cdots \exp[\frac{-i}{\hbar}\int^{t_1}_{t_0} H(\tau) d\tau] \Psi
(t_0)
\end{equation}
or
\begin{equation}
 \Psi (t)=\exp[\frac{-i}{\hbar}\int^{t_j}_{t_{j-1}} H(\tau) d\tau]
 \Psi(t_{j-1}), \hspace*{12mm} j=1,2,\cdots
\end{equation}
We then explore the possibility of replacing the Hamiltonian
$H(t)$ by its stepwise time varying approximation $\hat{H}(t)$
[14-15] which is defined as
\begin{equation}
 \hat{H}_j=\frac{1}{\Delta t_j}\int_{t_{j-1}}^{t_j} H (\tau)
 d\tau, \hspace*{12mm} t_{j-1} < t < t_j
\end{equation}
or
\begin{displaymath}
 \hat{H}_j \equiv H(\hat{t}_j), \hspace*{12mm} \mbox{with}
 \hspace* {6mm} \hat{t}_j=\frac{t_{j-1}+t_j}{2}.
\end{displaymath}
Now,  during each of the time intervals expressed above, the newly
defined Hamiltonian $\hat{H}_j$ is independent of time and the
typical intermediate state  becomes
\begin{equation}
 \Psi (t_j) \approx
  \exp[\frac{-i}{\hbar} \hat{H}_j \Delta t_j ]
 \Psi (t_{j-1})
\end{equation}
which implies
\begin{displaymath}
 \Psi (t)=\lim_{N\rightarrow \infty} \exp[\frac{-i}{\hbar} \hat{H}_N \Delta t_N
 ]\cdots \exp[\frac{-i}{\hbar} \hat{H}_1 \Delta t_1 ] \Psi(t_0).
\end{displaymath}
For the Hilbert space related to $\hat{H}_j$ (each $\hat{H}_j$
defines a Hilbert space and there are $N$ Hilbert spaces), we have
\begin{displaymath}
 \hat{H}_j \Psi_n^j (r) = E_n^j \Psi_n^j (r),
\end{displaymath}
where $E^j_n$ and $\Psi^j_n$ are the $n$th-eigenenergy and $n$th
normalized eigenfunction during $t_j < t < t_{j-1}$. Once the
wavefunction $\Psi (t_{j-1})$ is known, the wavefunction $\Psi
(t_j)$ can be expressed by
\begin{equation}
 \Psi (t_j) =
  \sum C_n^j \exp[\frac{-i}{\hbar} E_n^j \Delta t_j ]
 \Psi_n^j (r)
\end{equation}
where $C^j_n$ is determined by a projection
\begin{equation}
 C_n^j = \int \Psi (t_{j-1}) W_n^{j*} (r) dr.
\end{equation}
Using the Dirac notation, above expressions become
\begin{displaymath}
 \Psi (t_j) \rangle =
  \sum_n \exp[{-i} \omega_n^j  \Delta t_j ] |t_j,n\rangle
  \langle t_j,n|
 \Psi (t_{j-1})\rangle,  \hspace*{12mm} \omega_n^j=E_n^j/\hbar.
\end{displaymath}
It means for a short time interval a dynamical system and its
corresponding stationary system evolve in almost the same
way.\newline If the initial wavefunction $\Psi (t_{j-1}$ is
expanded in terms of the eigenfunctions during $\Delta t_{j-1}$
\begin{equation}
 \Psi (t_{j-1}) \rangle = \sum_{l} C_l |t_{j-1},l\rangle
\end{equation}
we then have
\begin{equation}
 \Psi (t_j) \rangle = \sum_{n} C_n |t_{j},n\rangle
\end{equation}
where
\begin{displaymath}
 C_n=\exp[{-i} \omega_n^j  \Delta t_j ] \sum_l C_l \langle
 t_j,n|t_{j-1},l\rangle.
\end{displaymath}
In fact we can have
\begin{equation}
 \Psi (t)\rangle=\sum_{k,\cdots,l,n} \exp[-i
 \theta_{k,\cdots,l,n}] |t_N,k\rangle\langle t_N,k|\cdots|t_2,l\rangle\langle t_2,l|
 |t_1,n\rangle\langle t_1,n| \Psi (t_0) \rangle,
\end{equation}
with
\begin{displaymath}
 \exp[-i
 \theta_{k,\cdots,l,n}]=\exp[-i(\omega_k^N \Delta_N
 +\cdots+\omega_l^2 \Delta t_2+\omega_n^1 \Delta_1)].
\end{displaymath}
Above results also mean that the multi-projection component $|t_N,
k\rangle$$\cdots$ $\langle$$t_2$$, l|t_1, n\rangle \langle t_1,
n|\Psi (t_0)\rangle$, as a part of the initial wavefunction,
indeed {\it passes through} the energy states labelled $n, l,
\cdots, k$ in the defined time-division sequence and should
eventually get the phase factor $\exp[-i
 \theta_{k,\cdots,l,n}]$ (the real Hamiltonian
is replaced by its stepwise time-varying counterparts; cf. [4-5]).
 Noting $\sum_n |t_j, n\rangle\langle t_j, n| \equiv 1$,
 we thus find an
interesting and important fact that if all phase factors of the
form $\exp(i\theta)$ disappeared from Eq. (9), the wavefunction
would not change at all. Note also that, as to above approach, it
is well known that the function $\exp[i f(t,r)] \Psi(t,r)$ where f
(t, r) is an arbitrary function of t and r, is also the system's
wavefunction provided that the gauge in question (the gauge choice
should be allowable in view of the fact that a certain gauge
transformation can always make the longitudinal vector field
vanish [14]) is allowed to be arbitrary. We remind the readers
that a phase factor of the form $\exp[i f(t,r)]$ can be a
nonuniformly continuous function and there are mathematical
complications in dealing with nonuniformly continuous function
[15], we are convinced that the gauge arbitrariness related to
$\exp[i f (t,r)] \Psi(t, r)$ should be excluded. \newline
The objective of above presentation is to illustrate that these
principles are indeed workable in terms of solving the
time-dependent Schr\"{o}odinger equation and to show that the
evolution of a quantum system can be characterized almost entirely
by phase dynamics.
\section{Results and Discussion}
We are now to consider the relevant issue related to the simple
knots [18-19] which is the our main focus here. The main
assumptions are : The final Hamiltonian is the same as the initial
Hamiltonian, the perturbation is relatively small and the action
time of the perturbation is relatively short (above approach is
similar to the path-integral approach as mentioned already and is
related to the phase dynamics of quantum mechanics). In
particular, considering this case : a harmonic oscillator
disturbed for a time $0<t<T$ and having the Hamiltonian
\begin{displaymath}
 H=\frac{p^2}{2}+x^2 \frac{S(t)}{2},
\end{displaymath}
where $S(t)$ is a stepwise function : $S=1$ for $t\le 0$,
$S(t)=\chi
>0$ for $0<t<T$, $S=1$ for $t\ge T$. We
can then obtain the system's wavefunction during $0<t<T$ based on
above approach
\begin{displaymath}
 \Psi (t)= \sum_n e^{-i E_{n,\chi} t} |n,\chi\rangle \langle n,\chi
 |0\rangle,
\end{displaymath}
where $E_{n,\chi}=\omega_{\chi} (n+1/2)$,
$\omega_{\chi}=\sqrt{\chi}$ ($\hbar=1$). Meanwhile after (time)
$t=T$,
\begin{displaymath}
 \Psi (t)=\sum_{n,m} \exp(-i E_{n,\chi}T)\exp[-i
 E_{m,1}(t-T)]|m,1\rangle\langle m,1|n,\chi\rangle\langle
 n,\chi|0\rangle.
\end{displaymath}
Now, an interesting observation occurs, considering the phase
character of this evolving system, once we assume that the system
is initially in the ground state and let $T$ be equal to
$4\pi/\omega_{\chi}$. This system will return the same ground
eigenstate after $t=T$ (due to $\exp({-i E_{n,\chi} T})=1$ and
$\sum_n |n,\chi\rangle \langle n,\chi | =1$; cf. equations above)
\begin{displaymath}
 \Psi (t)= \sum_m e^{-i E_{m,1} (t-T)} |m,1\rangle \langle m,1
 |0\rangle,
\end{displaymath}
and the only effect of the disturbance for the additional phase
factor could be tuned by  choices of $\chi$. To be specific, the
system seems to be completely frozen during $0 < t < T$ (this
resembles the phenomena predicted by the quantum Zeno effect; cf.,
e.g., [16-17]).
\newline  With above results and $\omega_{\chi}=1$ we then have
 for the
case of simple knot : Trefoil [18-19] as schematically shown in
Fig. 1. A trefoil is the simplest example of torus-knots, which
are obtained by winding around a torus in both directions
[10,12,17,20]. We can observe the evolution of certain system
along the trefoil-like trajectory during $0<t<T$ with respect to
any initial site or state, say, located at U or R or L (cf. Fig.
1). The system evolves starting from U, (as $T=4\pi$ for this kind
of trefoil-like paths traversing w.r.t. the vertex U), after
$t=T$, the system will return to U again. The above mentioned
results could thus be applied to this kind of path-traversing
($T=4\pi$) for any topological changes of trefoil-type knots
[7,18-19].
\newline Similarly, the $8_{18}$ knot (cf., e.g., [21]) can also be realized by selecting
$\omega_{\chi}=2/3$ ($T=6 \pi$; say, starting from the vertex K).
The illustration is shown in Fig. 2. To give a brief summary, we
already illustrated  a  possible traversing along the path of
trefoil-type and $8_{18}$ knots during  a characteristic time
period by considering a quantum-mechanic system which satisfies a
specific kind of phase dynamics of quantum mechanics. We believe
this presentation is relevant to the composite particle which is
present in the initial or final configuration. Other complicated
knot-type evolutions [22-26], the {\it unparticle} stuff [27] for
the specific quantum-mechanic system introduced above as well as
the role of Chern-Simons classes [28-29] based on the present
application will be investigated in the future. {\it
Acknowledgements.} {\small The author thanks the hospitality of
Chern (S.-S.) Inst. of Math. (Nankai Univ.) when the author
visited there around the mid. of 2008-Jan.}

\vspace{15mm}

\setlength{\unitlength}{1.00mm}   
\begin{picture}(120,44)(0,-6)
\thicklines \put(51,30){\oval(14.2,15)[br]}
\put(57.8,29.2){\oval(15,14)[bl]}
{\bezier{40}(58,30)(58,38.6)(53.5,38.6)}
{\bezier{40}(50,30.5)(50,38.6)(54.5,38.6)}
\put(66.6,26){\circle*{1}}
\put(41.4,26){\circle*{1}}
\put(54,38.6){\circle*{1}}
{\bezier{40}(58,22.5)(66.6,22.5)(66.6,26)}
{\bezier{40}(54,30)(66.6,30)(66.6,26)}
{\bezier{40}(50,22.5)(41.4,22.5)(41.4,26)}
{\bezier{40}(54,30)(41.4,30)(41.4,26)} \thinlines
\put(54,39.5){\makebox(0,0)[b]{\small U}}
\put(68.5,27){\makebox(0,0)[b]{\small R}}
\put(40,26.5){\makebox(0,0)[b]{\small L}}
\put(54,25.7){\makebox(0,0)[b]{\small o}}
\put(62,26){\makebox(0,0)[b]{\tiny x}}
\put(46,26){\makebox(0,0)[b]{\tiny x}}
\put(54,34){\makebox(0,0)[b]{\tiny x}}
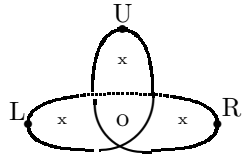
\put(10,2){\makebox(0,0)[bl]{\small Fig. 1 \hspace*{1mm} Schematic
(diagram) of a {\it loosely} hard trefoil [7,18-19]. Those points
of solid circle : }} \put(10,-2){\makebox(0,0)[bl]{\small U, R, L
could be prescribed as {\it vertices} or {\it lattice sites}.}}
\end{picture}

\vspace{10mm}

\setlength{\unitlength}{1.20mm}   
\begin{picture}(100,45)(0,-5)
\thicklines \put(40.8,30){\oval(14.5,15)[br]}
\put(47.9,29.5){\oval(15.5,14.2)[bl]}
{\bezier{35}(48,30)(48,38.6)(43.8,38.6)}
{\bezier{35}(40,30.5)(40,38.6)(44.2,38.6)}
\put(40.6,22.9){\oval(15.2,14.2)[tr]}
\put(47.7,22.5){\oval(15,15)[tl]}
{\bezier{35}(48,21.8)(48,13.9)(43.8,13.9)}
{\bezier{35}(40,22.5)(40,13.9)(44.2,13.9)}
{\bezier{35}(48,22.5)(56.6,22.5)(56.6,26)}
{\bezier{35}(48.7,30)(56.6,30)(56.6,26)}
{\bezier{35}(39.2,22.5)(31.4,22.5)(31.4,26)}
{\bezier{35}(50.2,30)(31.4,30)(31.4,26)} \thinlines
\put(43.8,22){\makebox(0,0){\tiny D}}
\put(44,12.2){\makebox(0,0){\tiny I}}
\put(44,39.8){\makebox(0,0){\tiny K}}
\put(30.2,26.2){\makebox(0,0){\tiny L}}
\put(44,30.6){\makebox(0,0){\tiny B}}
\put(49.5,31.5){\makebox(0,0){\tiny F}}
\put(38.8,31.5){\makebox(0,0){\tiny G}}
\put(49.5,20.5){\makebox(0,0){\tiny E}}
\put(48.8,26.2){\makebox(0,0){\tiny A}}
\put(38.8,20.6){\makebox(0,0){\tiny H}}
\put(39.4,26.2){\makebox(0,0){\tiny C}}
\put(58,26.2){\makebox(0,0){\tiny J}}
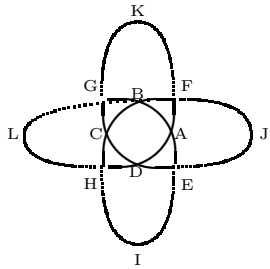
\put(12,1){\makebox(0,0)[bl]{\small Fig. 2 \hspace*{1mm} Schematic
(diagram) of a {\it loosely} hard $8_{18}$-knot [21].}}
\end{picture}

\end{document}